\documentclass[preprint2]{aastex}

\newcommand{\hii}{{\sc H~ii}\/ }

\def\simlt{\lower.5ex\hbox{$\; \buildrel < \over \sim \;$}}
\def\simgt{\lower.5ex\hbox{$\; \buildrel > \over \sim \;$}}
\def\arcdeg{\hbox{$^\circ$}}
\def\arcmin{\hbox{$^\prime$}}
\def\arcsec{\hbox{$^{\prime\prime}$}}
\slugcomment{Not to appear in Nonlearned J., 45. \today}

\begin{document}

\title{Physical Parameters of the Old Open Cluster Trumpler 5}
\author{Sang Chul Kim$^1$ and Hwankyung Sung$^2$}
\affil{$^1$ Korea Astronomy Observatory, Taejon 305-348, Republic of Korea;
   sckim@kao.re.kr \\
$^2$ Sejong University, Seoul 143-747, Republic of Korea; sungh@arcsec.sejong.ac.kr 
\\ 
...... \\
Accepted by J. of the Korean Astron. Soc. (2003 March issue)}

\shorttitle{Trumpler 5}
\shortauthors{Kim \& Sung}

\begin{abstract}
We present a study of the old open cluster Trumpler 5 (Tr 5), 
based on the CDS archival data.
  From the color-magnitude diagrams of Tr 5, we have found the positions 
of main-sequence turn-off (MSTO) and red giant clump (RGC) stars.
  Using the mean magnitude of the RGC stars, we have estimated the reddening
toward Tr 5, $E(B-V) = 0.60 \pm 0.10$.
Using the stars common in two data sets and the theoretical isochrones of Padova group, 
we have estimated the distance modulus $V_0-M_V=12.64 \pm 0.20$ (d $= 3.4 \pm 0.3$ kpc),
the metallicity [Fe/H] $= -0.30 \pm 0.10$, and the age of $2.4 \pm 0.2$ Gyr 
(${\rm log} ~t=9.38$).
  These metallicity and distance values are consistent with the relation
between the metallicity and the Galactocentric distance of other old
open clusters, for which we obtain the slope of $\Delta$[Fe/H]/$R_{gc}=
-0.064 \pm 0.010$ dex kpc$^{-1}$.
\end{abstract}

\keywords{open clusters and associations: individual (Trumpler 5) --
Galaxy: disk -- Galaxy: stellar content -- Galaxy: structure -- 
Hertzsprung-Russell diagram}

\section{Introduction}
Old open clusters, with ages greater than $\sim 1$ Gyr, 
are an important tool for the study of the formation and 
the early evolution of the Galactic disk (Friel 1995; 
Chen, Hou, \& Wang 2003).
Trumpler 5 (Tr 5; $\alpha$ = 06$^h$ 36$^m$ 32$^s$, $\delta$
= +09\arcdeg~ 29\arcmin~ 25\arcsec, J2000.0)
is located in anti-Galactic center region with very small galactic latitude
($l$ = 202.\arcdeg86, $b$ = +01.\arcdeg05).

  Dow \& Hawarden (1970) obtained distance modulus $(m-M)_V=14.40$ and
reddening $E(B-V)=0.80$ which give a true distance modulus of
$(m-M)_0=11.92$ (d=2.4 kpc) using total-to-selective extinction ratio of $R=3.1$, 
though their color-magnitude diagram have some scatter (Janes \& Adler 1982).
  Kalinowski (1974) and Kalinowski et al. (1974)
observed a very red ($B-V \approx $ 6.1 mag) 
carbon star, V493 Mon, and suggested that this star is possibly a member of Tr 5 
adopting $(m-M)_0 = 12.3\pm0.1$ (d=$2.9\pm0.1$ kpc) and $E(B-V)=0.48$ for Tr 5.
  Kalinowski (1975) performed photoelectric and photographic photometry
for 1300 stars in Tr 5, and presented the visual absorption A$_V \approx 2.1$ mag and 
diameter of \simgt 12\arcmin (8.4 pc) for Tr 5 at an assumed true distance 
modulus of $(m-M)_0 = 11.9\pm0.3$ (d=$2.4\pm0.3$ kpc).
  Piccirillo, Kalinowski, \& Wing (1977) found three M giants, three K giants,
and another carbon star with a color temperature of about 4000 K.
  These authors adopted $(m-M)_0 = 11.4$ (d=1.9 kpc) and $E(B-V)=0.6$ for Tr 5.
  Kalinowski (1979) have used $(m-M)_V=12.00$ and $E(B-V)=0.64$, which gives 
$(m-M)_0=10.02$ (d=1.0 kpc) using $R=3.1$ (Janes \& Adler 1982).
  Recently, Kaluzny (1998) have obtained $BVI_C$ CCD photometry of Tr 5
using Kitt Peak National Observatory 2.1m and 0.9m telescopes.
 He has obtained the parameters for Tr 5 as $(m-M)_0=12.4$ (d=3.0 kpc), 
$E(B-V)=0.58$, and age of 4.1 Gyr (${\rm log} ~t=9.61$).  
However, Kaluzny obtained the reddening and age of Tr 5 by comparing 
the photometric values of Tr 5 and M67 (Montgomery et al. 1993).

In this paper we present a new analysis of the data for Tr 5 
obtained by Kaluzny (1998).
Section 2 describes the data sets used in this study and 
Section 3 presents the color-magnitude diagrams of Tr 5.
In Sections 4 and 5, the physical parameters of Tr 5, such as reddening,
distance, metallicity and age, are derived.
Section 6 discusses the comparison of the derived results with those of 
Kaluzny (1998) and the metallicity-Galactocentric radius relation.
Finally, a summary and conclusions are given in Secion 7.

\section{Observational Material}
J. Kaluzny performed the observations of Tr 5 at the Kitt Peak National
Observatory on 1990 Nov. 11 (run \#1; KPNO 2.1m telescope + TEK1 CCD),
on 1991 Oct. 6 (run \#2; KPNO 2.1m telescope + T1KA CCD), and
on 1991 Oct. 10 -- 11 (run \# 3; KPNO 0.9m telescope + T1KA CCD).
The details on the observations and data reduction can be found in Kaluzny (1998).
Figure 1 shows the fields covered by the three data sets 
superposed on the Digitized Sky Survey image of Tr 5.
The non-rectangular shape of the field \#3 shows that
the photometry data of the field \#3 in CDS does not contain 
all the data of the stars in the rectangular field.

The data sets from the three observing runs were calibrated independently,
giving systematic differences less than 0.02 mag.
The photometric errors in the magnitude from the CDS data given 
by Kaluzny (1998) are plotted against the $V$ magnitude in Figure 2.

\section{Color-Magnitude Diagrams of Tr 5}
  Figure 3 shows the $V-(V-I)$ and $V-(B-V)$ color-magnitude diagrams (CMDs)
of the measured stars in Tr 5.
Upper panels are for the stars of good photometric qualities
and lower panels are for all the observed stars.
Panels (a) and (d) are for observing run \#1,
(b) and (e) are for the run \#2, and 
(c) and (f) are for the run \#3.

  The distinguishable features seen in the CMDs of Figure 3 are:
(i) there is a well-defined main-sequence, the top of which is probably
  located at $(B-V) \approx 1.1$, $(V-I) \approx 1.35$, and $V \approx 16.7$ mag
  (see Figure 2 of Kaluzny (1994) for the definitions of main-sequence turn-off (MSTO));
(ii) there is a well-developed red giant branch and red giant clump (RGC) stars
at $(B-V)_{RGC} = 1.55$, $(V-I)_{RGC} = 1.82$, and 
$V_{RGC} = 15.10$ mag, which is noted by crosses in the figures;
(iii) as it goes from panel (a) to (b) and to (c), the total number of stars
increases, the thickness of the main-sequence stars increases and 
there exist more stars (probably field stars) brighter than the MSTO at $V
\sim 16.7$ mag, which is reasonable and expectable since the observing run
\#1 covers the smallest area and the run \#3 covers the widest area.

\section{Reddening}
Since Tr 5 is located at low Galactic latitude ($b =$ +01.\arcdeg05),
it is expectable that the interstellar reddening toward Tr 5 could be
significant. We have estimated the reddening toward Tr 5 
using the mean color of the RGC stars.
Janes \& Phelps (1994) estimated the mean color and magnitude 
of the RGC in old open clusters to be
$(B-V)_{0, RGC} = 0.95 \pm 0.10$, and $M_{V, RGC} = 0.90 \pm 0.40$,
when the $V$ magnitude difference between the RGC and
the main-sequence turn-off of the clusters, $\delta V$, is greater than one.
  The mean colors of the RGC are estimated to be
$(B-V)_{RGC} = 1.55 \pm 0.01$ and $(V-I)_{RGC} = 1.82 \pm 0.01$, and
the corresponding mean magnitude is $V_{RGC} = 15.10 \pm 0.05$ mag.
  $\delta V$ is estimated to be $1.6 \pm 0.2$ mag, and
the resulting reddening value is estimated to be $E(B-V) = 0.60 \pm 0.10$.

\section{Distance, Metallicity and Age}
  Using the mean magnitude of the RGC (Janes \& Phelps 1994) 
and that of the RGC of Tr 5 obtained in the previous section,
we have estimated the distance modulus of Tr 5 as $V-M_V = 14.2 \pm 0.1$.

For the stars common in the data sets of \#1 and \# 2,
we have plotted the color-magnitude diagrams in Figure 4
and superimposed the ZAMS relation (Sung \& Bessell 1999; Sung 2001) and
the theoretical isochrones of Padova group (Bertelli et al. 1994) which are
reddened by $E(B-V)=0.60$ and $E(V-I)=0.80$ and shifted
according to the apparent distance modulus of $V-M_V=14.6$ (panels (a) and (b))
and $V-M_V=14.2$ (panels (c) and (d)). We have used the metallicity of
[Fe/H] $= -0.30$ and the age of 2.4 Gyr for Tr 5.
The distance modulus of $V-M_V=14.6$ gives better fit to the
main-sequence and sub-giant branch than that of $V-M_V = 14.2$.
Therefore, we take $V-M_V = 14.6 \pm 0.2$ as our final value.
Using the reddening values of $E(B-V)=0.60$ and $E(V-I)=0.80$ obtained above
and the total-to-selective extiontion ratio (Guetter \& Vrba 1989),
\begin{equation}
R_V = \frac{A_V}{E(B-V)} = 2.45 \frac{E(V-I)}{E(B-V)},
\end{equation}
we get $R_V = 3.26$, $A_V = 1.96$ mag, and the true distance modulus
of $V_0 - M_V = 12.64 \pm 0.2$ mag (d $= 3.4 \pm 0.3$ kpc). 

  We also have estimated the cluster age using 
the morphological age index (MAI) which was introduced by
Phelps, Janes, \& Montgomery (1994).  
We used the relation between the MAI and $\delta V$ given by
Phelps et al. (1994) and Janes \& Phelps (1994),
\begin{equation}
{\rm MAI}[{\rm Gyr}] = 0.73 \times 10^{(0.256 \delta V + 0.0662 \delta V^2)},
\end{equation}
where $\delta V$ represents the $V$ magnitude difference between 
the RGC and the MSTO of the clusters.
  From the value of $\delta V$ derived in Section IV, 
$1.6 \pm 0.2$ mag, we obtain a value for the age, MAI $= 2.8 \pm 0.6$ Gyr.
Even though the MAI was only intended to provide a relative age ranking
of clusters,
this age is consistent with the value obtained above from the isochrone fitting
within the error range.
  Figure 5 shows the Padova isochrone fitting with three age values of 
2.0 (left dotted lines), 2.4 (solid lines), and 2.8 (right dotted lines) Gyr.  
Finally, we take the age and the metallicity of Tr 5 as 
$2.4 \pm 0.2$ Gyr and [Fe/H] $= -0.30 \pm 0.10$ dex, respectively.

\section{Discussion}
\subsection{Comparison with the Results of Kaluzny}
  Kaluzny (1998) has obtained the age of Tr 5 to be $\sim 4.1$ Gyr 
by estimating the $\Delta (B-V)$ and $\Delta (V-I)$ (the difference 
in color between the cluster's MSTO and the red giant branch at 
the level of the clump).
  But the difficulty in estimating the position of the MSTO, as he stated, 
as well as the locus of red giant branch and the position of the RGC itself,
gives difference in the obtained ages between that of Kaluzny and ours
(2.4 Gyr).
  Kaluzny also obtained the reddening values of Tr 5, $E(B-V)=0.58$ and 
$E(V-I)=0.765$, by comparing the colors of the RGCs of Tr 5 and M67
assuming the age of 4.1 Gyr and metallicity the same as that of M67
for Tr 5.
  He derived the distance of Tr 5 as $(m-M)_0 \approx 12.4$ by
comparing the $I$-band luminosity of the 
RGC of Tr 5 and the absolute $I$-band luminosity of red clump stars
in the solar vicinity obtained from the Hipparcos distances ($M_I=-0.26$).

  In Figure 6 we showed the Padova isochrone fittings using 
the parameters of Tr 5 derived by Kaluzny (1998) : age=4.1 Gyr, 
[Fe/H]=0.00 dex,
$E(B-V)=0.58$, $E(V-I)=0.765$, $A_V = 2.947 \times E(B-V) = 1.71$,
and $V-M_V = 12.4 + A_V = 14.11$,
on (a) the $V-(B-V)$ color-magnitude diagram for 
the stars commonly contained in the photometry files of observing runs \#1 and \#2,
and color-magnitude diagrams for the stars of observing runs \#1, \#2, and
\#3 in panels (b), (c), and (d), respectively.
It is clearly seen that the parameters given by Kaluzny do not fit well
the color-magnitude diagrams of observed stars.

Figure 7 shows the Padova isochrone fittings using the parameters
of Tr 5 derived in this study: age=2.4 Gyr, [Fe/H]=$-0.30$ dex,
$E(B-V)=0.60$, $E(V-I)=0.80$, $A_V = 3.26 \times E(B-V) = 1.96$,
and $V-M_V = 12.64 + A_V = 14.60$,
on (a) the $V-(B-V)$ color-magnitude diagram for
the stars commonly contained in the photometry files of the observing runs \#1 and \#2,
and color-magnitude diagrams for the stars of the observing runs \#1, \#2, and
\#3 in panels (b), (c), and (d), respectively.
Comparison of Figures 6 and 7 shows that the parameters derived in 
this study better fits the observed color-magnitude diagrams.
Since the ratio of field stars and binary stars to the member stars of Tr 5
might increases as we go from panels (a) to (d),
the good fitting of the isochrone in Figure 7 (a) (and (b))
is noteworthy among the four panels.

\subsection{[Fe/H]--Galactocentric Radius Relation}
  It is known that open clusters that are mainly in the Galactic disk 
follow the negative radial gradient of metallicity 
just as \hii regions, bright blue stars, red giants, 
and planetary nebulae (Portinari \& Chiosi 1999;
Hou, Prantzos, \& Boissier 2000).
  From the studies of open clusters, various authors have presented 
the slope of the Galactocentric radial [Fe/H] gradient 
($\Delta$[Fe/H]/$R_{gc}$) : 
$-0.091 \pm 0.014$ dex kpc$^{-1}$ (Friel 1995),
$-0.09$ dex kpc$^{-1}$ (Carraro, Ng, \& Portinari 1998),
$-0.086 \pm 0.011$ dex kpc$^{-1}$ (Park \& Lee 1999), and
$-0.06 \pm 0.01$ dex kpc$^{-1}$ (Friel 1999).
  Based on updated abundance calibration of spectroscopic indices measuring
Fe and Fe-peak element blends on 39 old open clusters, 
Friel et al. (2002) have estimated the abundance gradient of $-0.06 \pm 0.01$ 
dex kpc$^{-1}$ over a range in the Galactocentric radii of 7 to 16 kpc.
From compilation of 119 open cluster catalogs, Chen, Hou, \& Wang (2003)
have estimated the gradient of $-0.063 \pm 0.008$ dex kpc$^{-1}$.

  Compiling the data of old open clusters given by Friel (1995), 
Wee \& Lee (1996), Kassis, Friel, \& Phelps (1996),
Lee (1997), Ann, Park, \& Kang (1998), and Friel et al. (2002),
we made our catalog file of 50 old open clusters.
We adopted the Galactocentric distance of the Sun as 8.5 kpc.
  Figure 8 shows the Galactocentric radial [Fe/H] gradient for the 
50 old open clusters.  
  Open square and filled square are the position of Tr 5 based
on the parameters given by Kaluzny (1998; $R_{gc}=11.32$ kpc) 
and this study ($R_{gc}=11.71$ kpc), respectively.
  The solid line is a least-squares fit to the data that yields an
[Fe/H] gradient of $\Delta$[Fe/H]/$R_{gc}=-0.064 \pm 0.010$ dex kpc$^{-1}$.
  This value is in good agreement with the values obtained by others,
especially with the values derived from recent studies 
and the position of Tr 5 in Figure 8 obtained by using the parameters derived
in this study is more consistent with the mean trend of the other
old open clusters.

\section{Summary and Conclusions}
  We have presented the analysis of the photometry of the old open
cluster Tr 5 using the CDS archival data obtained by Kaluzny.
  We have redetermined the reddening, distance, metallicity, and
age of this cluster and summarized them in Table 1.
  These parameters are more consistent with the 
Galactocentric radial [Fe/H] gradient from other old open 
clusters.

\begin{table*}[t]
\begin{center}
{\bf Table 1.}~~Basic Information of Trumpler 5 \\
\vskip 3mm
{\small
\setlength{\tabcolsep}{1.2mm}
\begin{tabular}{llc} \hline\hline
Parameter & Information & Reference \\
\hline
Other names & C0634+094, OCL 494, Lund 237, Collinder 105 & Lyng\r{a} 1987 \\ 
$\alpha_{2000}$, $\delta_{2000}$ & 06$^h$ 36$^m$ 32$^s$, +09\arcdeg~ 29\arcmin~ 25\arcsec &
	This study \\
$l$, $b$ & 202.\arcdeg86, +01.\arcdeg05 & Lyng\r{a} 1987 \\ 
Trumpler class & III 1 r & Lyng\r{a} 1987 \\ 
Reddening, $E(B-V)$ & $0.60 \pm 0.10$ mag & This study \\
Distance modulus, $V_0 - M_V$ & $12.64 \pm 0.2$ mag & This study \\
Distance, d  & $3.4 \pm 0.3$ kpc & This study \\
Galactocentric distance, $R_{gc}$ & $11.71 \pm 0.29$ kpc & This study \\
Metallicity, [Fe/H] & $-0.30 \pm 0.10$ dex & This study \\
Age, $t$ & $2.4 \pm 0.2$ Gyr (${\rm log} ~t=9.38$) & This study \\
\hline
\end{tabular}
} 
\end{center}
\end{table*}

\acknowledgements
SCK is grateful to In-Soo Yuk for providing his routines of the Padova isochrones.
The authors would like to thank the referee, Prof. H. B. Ann for his careful 
reading of the manuscript which helped to clarify several issues.
We employed catalogues from CDS/Simbad (Strasbourg) and Digitized Sky Survey
images from the Space Telescope Science Institute.
This work is the result of research activities (Astrophysical Research Center 
for the Structure and Evolution of the Cosmos - ARCSEC) supported by Korea 
Science \& Engineering Foundation (HS).


\vfill
\begin{figure} 
\figurenum{1}
\plotone{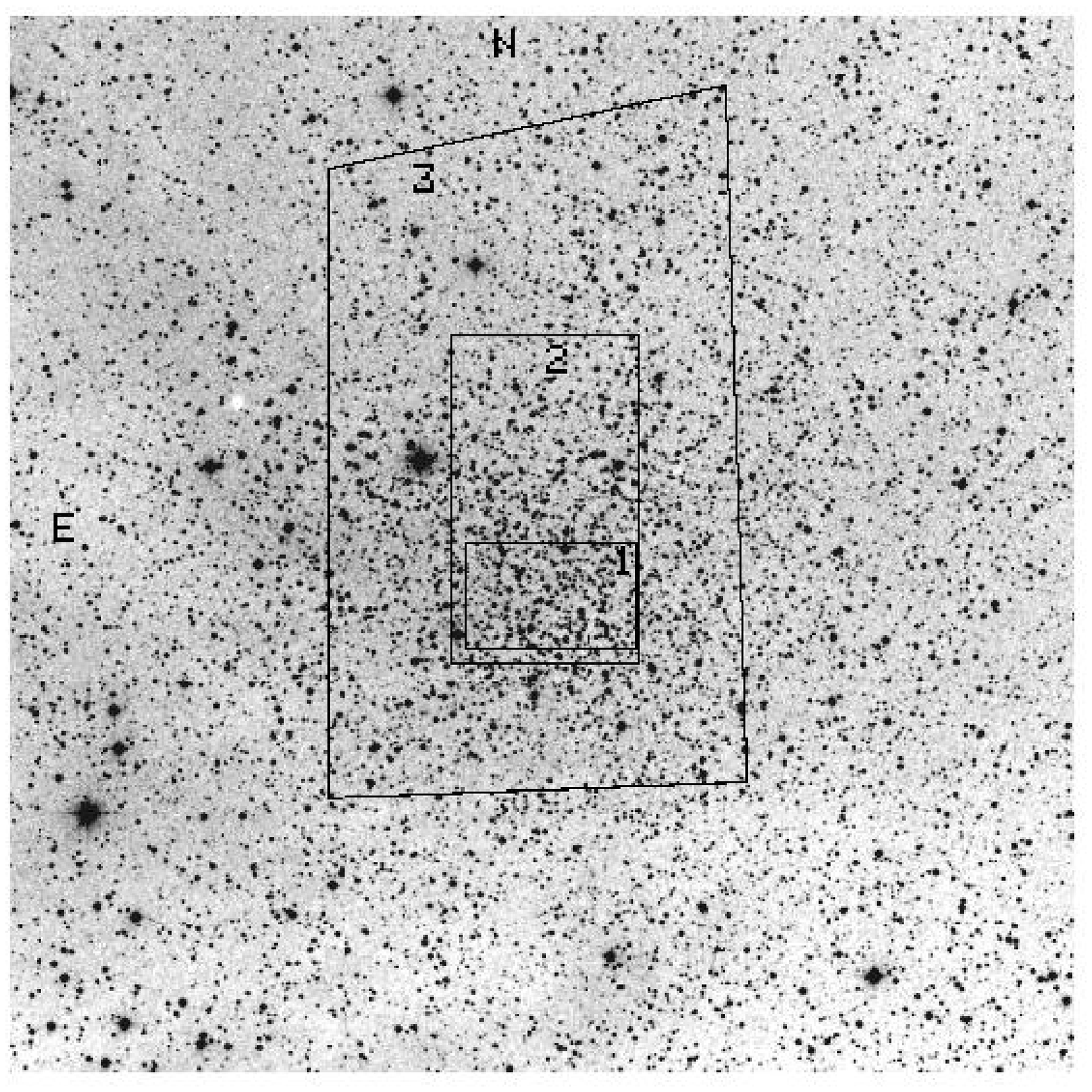} 
\figcaption{
The fields covered by the three observing runs
superposed on the Digitized Sky Survey image of Tr 5.
The center of the image is $\alpha_{2000}=06^h~ 36^m~ 32^s$,
$\delta_{2000}=$+09\arcdeg~ 29\arcmin~ 25\arcsec, and
the size of the field is 30\arcmin $\times$ 30\arcmin.
North is at the top and east is to the left.
Regions labeled ``1'', ``2'', and ``3'' represent, respectively,
the regions covered by the observing runs of \#1, \#2, and \#3.
See text for the details.
}
\end{figure}

\begin{figure}[2]
\figurenum{2}
\plotone{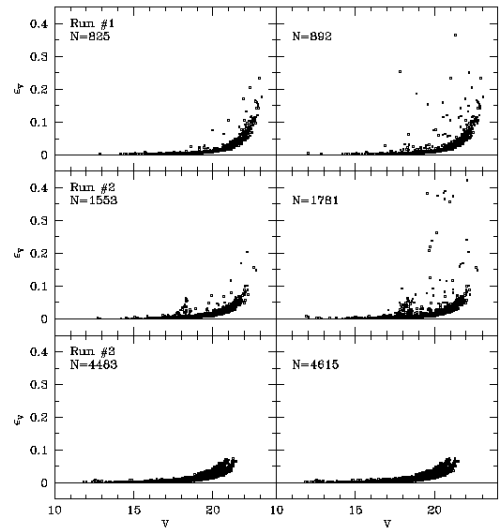}  
\figcaption{
Photometric errors from CDS data 
given by Kaluzny (1998) as a function of magnitude.
Top rows are for observing run \#1,
middle rows are for the run \#2, and bottom rows are for the run \#3.
Left panels are for stars of good photometric qualities
and right panels are for all the observed stars.
}
\end{figure}

\begin{figure}[3]
\figurenum{3}
\plotone{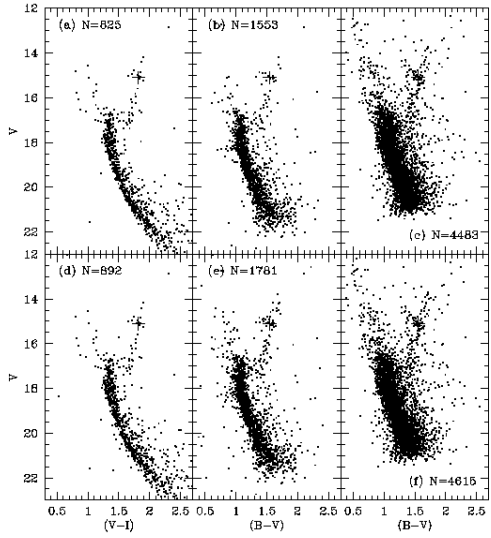}  
\figcaption{
Color-magnitude diagrams of Tr 5.
Upper panels are for stars of good photometric qualities
and lower panels are for all the observed stars.
Panesl (a) and (d) are for observing run \#1,
(b) and (e) are for the run \#2, and
(c) and (f) are for the run \#3.
The crosses represent the positions of the red giant clump (RGC).
}
\end{figure}

\begin{figure}[4]
\figurenum{4}
\plotone{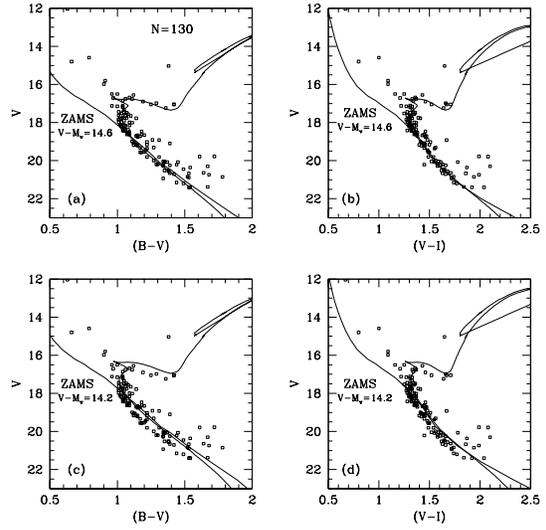}  
\figcaption{
$V-(B-V)$ (panels (a) and (c)) 
and $V-(V-I)$ (panels (b) and (d))
color-magnitude diagrams for the 130 stars commonly contained in
the photometry files of observing runs \#1 and \#2.
The solid lines represent
the ZAMS relation (Sung \& Bessell 1999, Sung 2001) and
the Padova isochrones (Bertelli et al. 1994), which are
reddened by $E(B-V)=0.60$ and $E(V-I)=0.80$ and shifted
according to the distance modulus of $V-M_V=14.6$ ((a) and (b))
and $V-M_V=14.2$ ((c) and (d)).
The distance modulus of $V-M_V=14.6$ gives better fit to the
main-sequence and sub-giant branch.
}
\end{figure}

\begin{figure}[5]
\figurenum{5}
\plotone{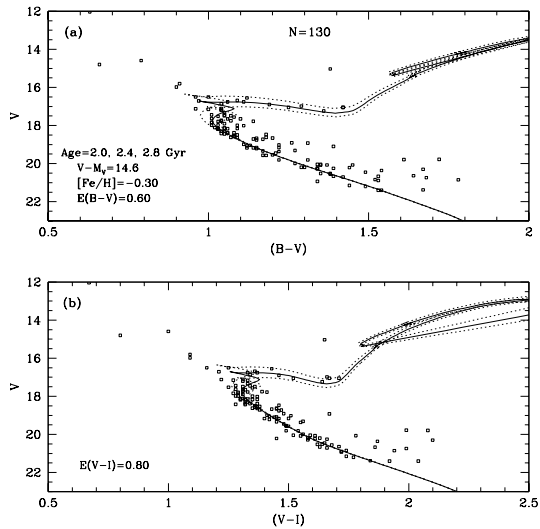}  
\figcaption{
The Padova isochrone fitting on 
the $V-(B-V)$ (a) and
$V-(V-I)$ (b) color-magnitude diagrams with three age values of
2.0 (left dotted lines), 2.4 (solid lines), and 2.8 (right dotted lines) Gyr.
}
\end{figure}

\begin{figure}[6]
\figurenum{6}
\plotone{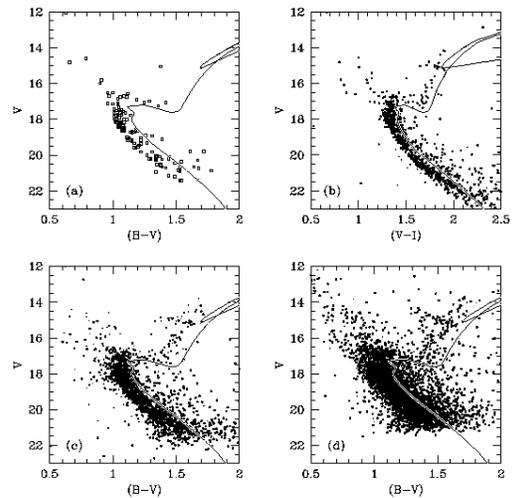}  
\figcaption{
The Padova isochrone fitting using
the parameters of Tr 5 derived by Kaluzny (1998) : age=4.1 Gyr, [Fe/H]=0.00,
$E(B-V)=0.58$, $E(V-I)=0.765$, $A_V = 2.947 \times E(B-V) = 1.71$,
and $V-M_V = 12.4 + A_V = 14.11$,
on (a) the $V-(B-V)$ color-magnitude diagram for
the stars commonly contained in the photometry files of the observing runs \#1 and \#2,
(b) the $V-(V-I)$ color-magnitude diagram for the stars of the observing run \#1,
(c) the $V-(B-V)$ color-magnitude diagram for the stars of the observing run \#2, and
(d) the $V-(B-V)$ color-magnitude diagram for the stars of the observing run \#3.
}
\end{figure}

\begin{figure}[7]
\figurenum{7}
\plotone{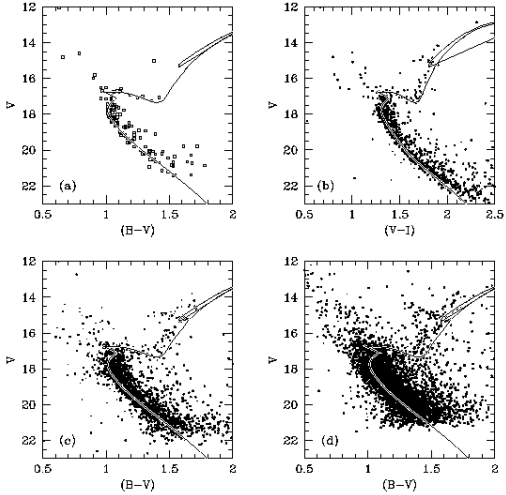}  
\figcaption{
The Padova isochrone fitting using
the parameters of Tr 5 derived in this study : age=2.4 Gyr, [Fe/H]=$-0.30$,
$E(B-V)=0.60$, $E(V-I)=0.80$, $A_V = 3.26 \times E(B-V) = 1.96$,
and $V-M_V = 12.64 + A_V = 14.60$,
on (a) the $V-(B-V)$ color-magnitude diagram for
the stars commonly contained in the photometry files of the observing runs \#1 and \#2,
(b) the $V-(V-I)$ color-magnitude diagram for the stars of the observing run \#1,
(c) the $V-(B-V)$ color-magnitude diagram for the stars of the observing run \#2, and
(d) the $V-(B-V)$ color-magnitude diagram for the stars of the observing run \#3.
}
\end{figure}

\begin{figure}[8]
\figurenum{8}
\plotone{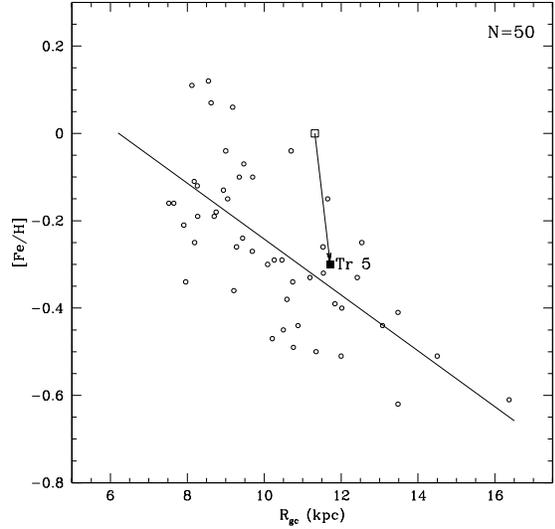}  
\figcaption{
Radial abundance gradient for 50 old open clusters.
The solid line is a least-squares fit to the data that yields an
abundance gradient of $\Delta$[Fe/H]/$R_{gc}=-0.064 \pm 0.010$ dex kpc$^{-1}$.
Open square and filled square are the position of Tr 5 based
on the parameters given by Kaluzny (1998) and this study, respectively.
}
\end{figure}

\end{document}